\documentclass[twocolumn,aps,pra,showpacs,groupedaddress,floatfix,letterpaper,superscriptaddress]{revtex4-2}
\usepackage{ae}
\usepackage[utf8]{inputenc}
\usepackage{graphicx}
\usepackage{hyperref}
\usepackage{amsfonts}
\usepackage{amsmath}
\usepackage{color}
\usepackage{braket}
\usepackage{enumerate}
\usepackage{xspace}
\usepackage[english]{babel}
\usepackage{widetable}

\def\lsim{\mathrel{\rlap{\lower3pt\hbox{$\sim$}}
    \raise2pt\hbox{$<$}}}
\def\gsim{\mathrel{\rlap{\lower3pt\hbox{$\sim$}}
    \raise2pt\hbox{$>$}}}

\def\ppbar{\mbox{(anti-)}proton\xspace}
\def\ppbars{\mbox{(anti-)}protons\xspace}
\def\ppbarss{\mbox{(anti-)}proton's\xspace}

\def\Ppbars{\mbox{(Anti-)}protons\xspace}
\def\Be{$^{9}$\text{Be}$^{+}$\xspace}
\def\Ca{$^{40}$\text{Ca}$^{+}$\xspace}

\usepackage{letltxmacro}
\LetLtxMacro{\ORIGselectlanguage}{\selectlanguage}
\makeatletter
\DeclareRobustCommand{\selectlanguage}[1]{%
	\@ifundefined{alias@\string#1}
	{\ORIGselectlanguage{#1}}
	{\begingroup\edef\x{\endgroup
			\noexpand\ORIGselectlanguage{\@nameuse{alias@#1}}}\x}%
}
\newcommand{\definelanguagealias}[2]{%
	\@namedef{alias@#1}{#2}%
}
\makeatother
\definelanguagealias{en}{english}
\definelanguagealias{En}{english}
\definelanguagealias{de}{german}
\definelanguagealias{EN}{english}


\begin{document}	
\title{Quantum logic inspired techniques for spacetime-symmetry tests with \ppbars}
\author{Juan M.\ Cornejo}
\affiliation{Institut für Quantenoptik, Leibniz Universität Hannover, Welfengarten 1, 30167 Hannover, Germany}
\affiliation{Laboratorium für Nano- und Quantenengineering, Leibniz Universität Hannover, Schneiderberg 39, 30167 Hannover, Germany}
\author{Ralf Lehnert}
\affiliation{Institut für Quantenoptik, Leibniz Universität Hannover, Welfengarten 1, 30167 Hannover, Germany}
\affiliation{Indiana University Center for Spacetime Symmetries, Bloomington, IN 47405, USA}
\author{Malte Niemann}
\affiliation{Institut für Quantenoptik, Leibniz Universität Hannover, Welfengarten 1, 30167 Hannover, Germany}
\affiliation{Laboratorium für Nano- und Quantenengineering, Leibniz Universität Hannover, Schneiderberg 39, 30167 Hannover, Germany}
\author{Johannes Mielke}  
\affiliation{Institut für Quantenoptik, Leibniz Universität Hannover, Welfengarten 1, 30167 Hannover, Germany}
\affiliation{Laboratorium für Nano- und Quantenengineering, Leibniz Universität Hannover, Schneiderberg 39, 30167 Hannover, Germany}
\author{Teresa Meiners}
\affiliation{Institut für Quantenoptik, Leibniz Universität Hannover, Welfengarten 1, 30167 Hannover, Germany}
\affiliation{Laboratorium für Nano- und Quantenengineering, Leibniz Universität Hannover, Schneiderberg 39, 30167 Hannover, Germany}
\author{Amado Bautista-Salvador}
\affiliation{Institut für Quantenoptik, Leibniz Universität Hannover, Welfengarten 1, 30167 Hannover, Germany}
\affiliation{Laboratorium für Nano- und Quantenengineering, Leibniz Universität Hannover, Schneiderberg 39, 30167 Hannover, Germany}
\affiliation{Physikalisch-Technische Bundesanstalt, Bundesallee 100, 38116 Braunschweig, Germany}
\author{Marius Schulte}
\affiliation{Institut für Theoretische Physik, Leibniz Universität Hannover, Appelstra{\ss}e 2, 30167 Hannover, Germany}
\author{Diana Nitzschke}
\affiliation{Institut für Theoretische Physik, Leibniz Universität Hannover, Appelstra{\ss}e 2, 30167 Hannover, Germany}
\author{Matthias J.\ Borchert}
\affiliation{Institut für Quantenoptik, Leibniz Universität Hannover, Welfengarten 1, 30167 Hannover, Germany}
\affiliation{Physikalisch-Technische Bundesanstalt, Bundesallee 100, 38116 Braunschweig, Germany}
\affiliation{Ulmer Fundamental Symmetries Laboratory, RIKEN, 2-1 Hirosawa, Wako, Saitama 351-0198, Japan}
\author{Klemens Hammerer}
\affiliation{Institut für Theoretische Physik, Leibniz Universität Hannover, Appelstra{\ss}e 2, 30167 Hannover, Germany}
\author{Stefan Ulmer}
\affiliation{Ulmer Fundamental Symmetries Laboratory, RIKEN, 2-1 Hirosawa, Wako, Saitama 351-0198, Japan}
\author{Christian Ospelkaus}         
\affiliation{Institut für Quantenoptik, Leibniz Universität Hannover, Welfengarten 1, 30167 Hannover, Germany}
\affiliation{Laboratorium für Nano- und Quantenengineering, Leibniz Universität Hannover, Schneiderberg 39, 30167 Hannover, Germany}
\affiliation{Physikalisch-Technische Bundesanstalt, Bundesallee 100, 38116 Braunschweig, Germany}

\begin{abstract}
	
Cosmological observations as well as theoretical approaches to physics beyond the Standard Model provide strong motivations for experimental tests of fundamental symmetries, such as CPT invariance. In this context, the availability of cold baryonic antimatter at CERN has opened an avenue for ultrahigh-precision comparisons of protons and antiprotons in Penning traps. This work discusses an experimental method inspired by quantum logic techniques that will improve particle localization and readout speed in such experiments. The method allows for sympathetic cooling of the \ppbar to its quantum-mechanical ground state as well as the readout of its spin alignment, replacing the commonly used continuous Stern-Gerlach effect. Both of these features are achieved through coupling to a laser-cooled `logic' ion co-trapped in a double-well potential. This technique will boost the measurement sampling rate and will thus provide results with lower statistical uncertainty, contributing to stringent searches for time dependent variations in the data. Such measurements ultimately yield extremely high sensitivities to CPT violating coefficients acting on baryons in the Standard-Model Extension, will allow the exploration of previously unmeasured types of symmetry violations, and will enable antimatter-based axion-like dark matter searches with improved mass resolution.

\end{abstract}

\maketitle

\section{Introduction}
\label{intro}

The study of symmetries and their breakdown is closely intertwined with the development of the Standard Model (SM) and continues to be a cornerstone of experimental and theoretical explorations aiming to uncover more fundamental physics. While most discrete spacetime symmetries are known to be violated in weak interactions, there is no credible experimental evidence of departures from the combined transformation of charge conjugation (C), parity inversion (P) and time reversal (T)~\cite{particle_data_group_review_2018,kostelecky_data_2011}. The theoretical underpinning of this result is provided by the CPT theorem~\cite{pauli_exclusion_1955}, which states that a realistic Lorentz-symmetric quantum field theory must also be CPT invariant. This profound result entails an exact symmetry between the properties of matter--antimatter conjugates.

However, the observable universe appears to exhibit an imbalance between baryons and antibaryons~\cite{canetti_matter_2012}. Although the SM contains the qualitative features for generating a baryon asymmetry, it struggles to explain the asymmetry quantitatively: the required amount of CP violation has thus far eluded solid experimental identification in various sectors of the SM~\cite{particle_data_group_review_2018,canetti_matter_2012}, and the large Higgs mass can impede the necessary departures from thermal equilibrium in the early universe~\cite{particle_data_group_review_2018,kajantie_electroweak_1996}. Planck-suppressed CPT breaking, on the other hand, may provide an alternative mechanism, as it can evade both of these conditions~\cite{bertolami_cpt_1997}. Moreover, a number of theoretical ideas involving physics beyond the SM can accommodate small deviations from CPT symmetry~\cite{page_is_1980,wald_quantum_1980,hawking_unpredictability_1982,alan_kostelecky_cpt_1991}. This situation has spurred the development of the Standard-Model Extension (SME) test framework for the theoretical description of such effects~\cite{colladay_lorentz-violating_1998} and has led to intensified efforts to perform matter--antimatter comparisons in general and improved CPT tests with baryons in particular~\cite{kostelecky_data_2011}.

The rapid progress and interest in ultrahigh-precision studies with baryons is evident from a number of recent measurements involving the \ppbar.
These include proton experiments performed by the BASE (Baryon-Antibaryon Symmetry Experiment) collaboration at Mainz~\cite{mooser_direct_2014,schneider_double-trap_2017} as well as measurements with both protons and antiprotons performed at CERN's Antiproton Decelerator (AD) by several collaborations. 
These efforts have yielded key measurements
involving charge-to-mass ratios and magnetic moments of these particles.

One of these was performed by the TRAP collaboration at CERN’s low-energy antiproton ring LEAR by cyclotron-frequency comparisons of antiprotons and protons~\cite{gabrielse_special_1995}, which was subsequently improved by using negatively-charged hydrogen ions~\cite{gabrielse_precision_1999}. Later, a more sensitive result 
for the proton-to-antiproton charge-to-mass ratio $(q/m)_{p}/(q/m)_{\bar{p}}= 1+1(69)\times 10^{-12}$ was obtained from cyclotron-frequency comparisons of a single antiproton and a H$^{-}$ ion by using the BASE apparatus at CERN's AD~\cite{ulmer_high-precision_2015}. BASE has also performed a 0.8 ppm~\cite{nagahama_sixfold_2017} measurement of the antiproton magnetic moment 
improving on a previous measurement~\cite{disciacca_resolving_2013} from the ATRAP collaboration
by a factor of six. 
Using single spin-flip resolution~\cite{smorra_observation_2017} 
and an advanced multi-trap method, 
BASE further improved this result
with the most precise determination of the antiproton magnetic moment 
with a fractional precision of 1.5 ppb in units of the nuclear magneton~\cite{smorra_parts-per-billion_2017}. 
This represents the first time that a measurement sensitivity 
for an antimatter property 
surpassed that of its matter conjugate.
This work has also placed constraints 
on CPT-odd SME coefficients 
at the $10^{-24}\;$GeV level~\cite{smorra_parts-per-billion_2017}, 
which is comparable in sensitivity 
to the corresponding results for the electron~\cite{van_dyck_new_1987,dehmelt_past_1999}, 
and outperforms limits extracted from muon $g$-factor measurements~\cite{muon_g-2_collaboration_final_2006,muon_g-2_collaboration_search_2008}. 
BASE-Mainz has subsequently continued to refine 
our knowledge of the proton magnetic moment 
at the sub-ppb\ level~\cite{schneider_double-trap_2017}.
The BASE measurements have adapted techniques originally developed by Dehmelt for electrons and positrons~\cite{van_dyck_new_1987,dehmelt_experiments_1990,dehmelt_past_1999}, combined them with the double-trap method~\cite{haffner_high-accuracy_2000}, and pushed their limits for the extremely challenging case of the \ppbar. 
This has enabled the measurement of the proton and antiproton magnetic moment at a respective relative precision of around 30 and 3000 times better 
than the previous best values 
performed by the hydogen MASER experiment~\cite{essen_frequency_1971} 
and the ATRAP collaboration ~\cite{disciacca_resolving_2013}, 
respectively. More recently, this has been used to tighten constraints on a possible axion-antiproton coupling, improving by up to five orders of magnitude the astrophysics-based previous measurements~\cite{smorra_direct_2019}. 

In parallel, a range of sophisticated laser-based techniques have been developed for the preparation, manipulation, and detection of atomic ions~\cite{wineland_experimental_1998-1,wineland_nobel_2013}. Early on, it was recognized that these techniques could also be used to support control and precision spectroscopy of other quantum systems~\cite{heinzen_quantum-limited_1990}, such as the \ppbar, ultimately leading to the experimental realization of quantum logic spectroscopy~\cite{schmidt_spectroscopy_2005}, the development of the single-ion aluminum clock~\cite{rosenband_observation_2007,chou_frequency_2010}, and more recently quantum logic spectroscopy of molecular ions~\cite{wolf_non-destructive_2016,chou_preparation_2017} and even highly charged ions~\cite{leopold_cryogenic_2019}. In Penning traps, applications to mass spectrometry have been proposed by using a single \Ca ion as a sensor in a double Penning trap system~\cite{rodriguez_quantum_2012}. More recently, the idea of using two Penning traps connected by a common electrode has been proposed to improve cooling in present $g$-factor experiments~\cite{bohman_sympathetic_2018}. In previous work~\cite{niemann_cpt_2013,smorra_base_2015}, we have already discussed concepts for coupling of atomic ions to single \ppbars using the double-well technique first suggested in Ref.~\cite{wineland_experimental_1998-1} and experimentally demonstrated in radio-frequency (rf) traps~\cite{brown_coupled_2011,harlander_trapped-ion_2011}. As a first experimental step, we have reported on the prospects of coupling pairs of \Be ions in a double-well potential and on the implementation of Doppler cooling of \Be ions in a specifically constructed 5\,T cryogenic Penning-trap setup~\cite{niemann_cryogenic_2019}. Finally, we have already discussed the realization of elementary quantum logic operations with (anti-)protons in Penning traps~\cite{nitzschke_elementary_2020}.

In this work, 
we discuss in detail the experimental conditions 
of a scenario for implementing quantum logic spectroscopy of \ppbars.
We also provide a few comments on the expected implications 
for constraints on SME coefficients. 

The outline of this paper is as follows. Section~\ref{Present status} reviews the main techniques currently employed in Penning-trap experiments. A brief review of the interplay between CPT and Lorentz violation, the SME test framework, and the prediction of sidereal variations is presented in Sec.~\ref{SME implications}. In Sec.~\ref{Experimental concept}, our experimental concept is presented in more detail. A time and error budget with projected constraints on various SME coefficients is presented in Sec.~\ref{Sensitivities and SME implications}. Finally, a short summary is contained in Sec.~\ref{Summary and outlook}. 

\section{Principles of \lowercase{\textit{\textbf{g}}}-factor measurements in Penning traps}
\label{Present status}

\begin{figure}[t]
	\centering
	\includegraphics[width=0.45\textwidth]{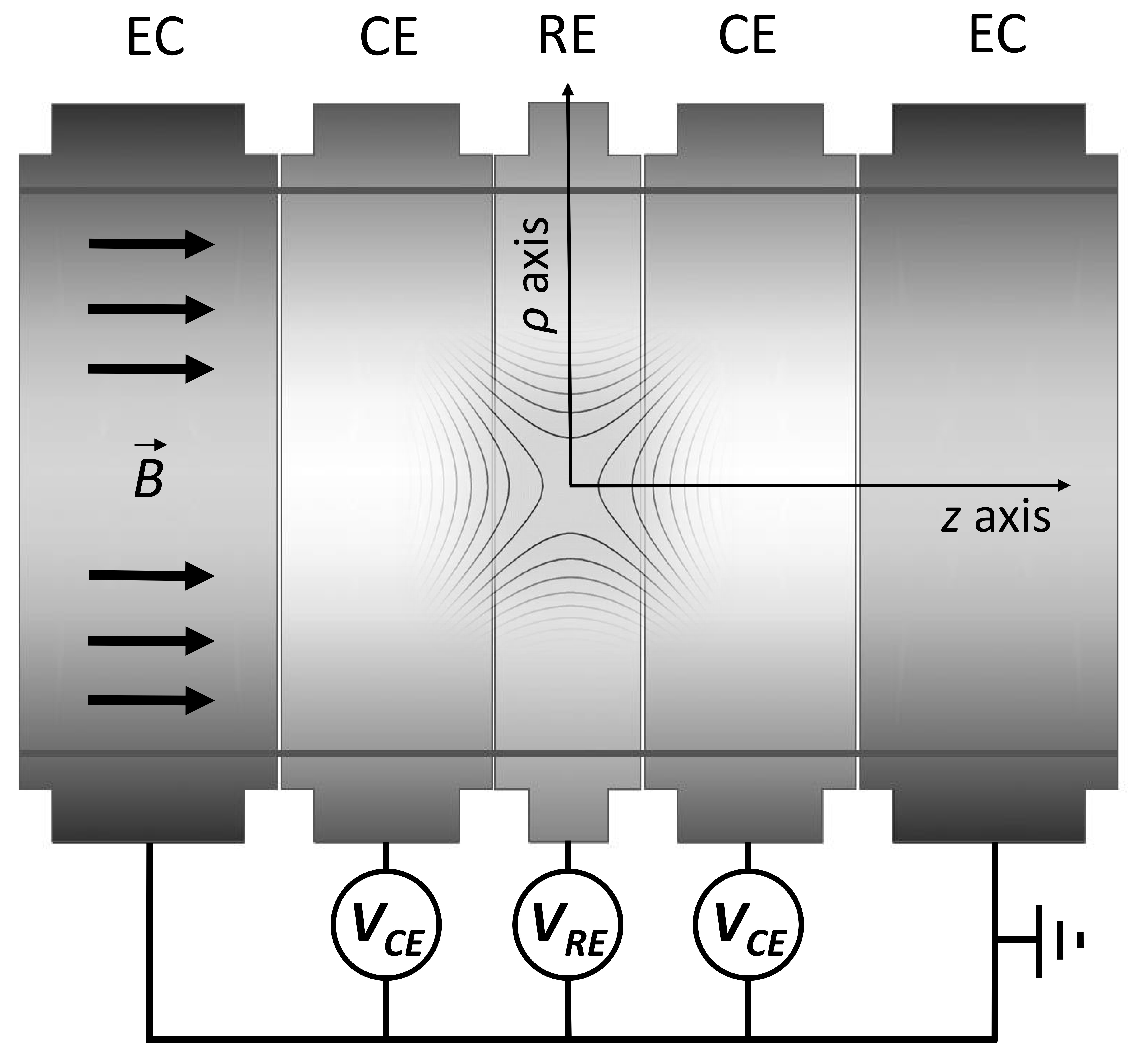}
	\caption{Schematic drawing of a Penning trap with cylindrical geometry. The electrostatic field is created by two constant voltages $V_{RE}$ and $V_{CE}$ applied to the ring electrode (RE) and correction electrodes (CE) with respect to the endcaps (EC), respectively. The equipotential lines are sketched in black around the center of the trap. An axial magnetic field $\vec{B}$ is superimposed for radial confinement.}
	\label{fig:trap}
\end{figure}

The Penning trap is an essential tool for high-precision measurements 
of fundamental properties of charged particles, 
such as charge-to-mass ratios or $g$-factors. 
In this device, a charged particle of interest is stored by superimposing an electric quadrupole field and a strong homogeneous magnetic field $\vec{B}$ in the axial direction. Figure~\ref{fig:trap} shows a sketch of a Penning trap with cylindrical shape where a quadrupole electric field is generated by applying a voltage $V_{RE}$ between the ring and endcap electrodes and a voltage $V_{CE}$ between the correction and endcap electrodes. By choosing a proper set of voltages as well as an optimized trap geometry, a harmonic potential shape along the trap axis $V(z)\approx V_{RE}\,C_{2}\,z^2$ is obtained, where $C_{2}$ is the unique nonzero coefficient of a multipole expansion~\cite{gabrielse_cylindrical_1984}.  

The trajectory of the ion in the trap can be described as a superposition of three motions, one in the axial direction and two in the radial plane, known as the modified cyclotron and magnetron motions. The axial motion has characteristic frequency 
\begin{equation}
\label{z_freq}
\nu_z=\frac{1}{2\pi}\sqrt{\frac{2\,q\,V_{RE}\,C_{2}}{m}},
\end{equation}
where $m$ and $q$ are the mass and charge of the stored particle, respectively. In the radial plane, the characteristic frequencies are given by
\begin{equation}
\label{rad_freq}
\nu_{\pm}=\frac{1}{2}\left(\nu_c\pm\sqrt{\nu_c^2-2\nu_z^2}\right),
\end{equation}
where $\nu_+$ and $\nu_-$ are the modified cyclotron and magnetron frequencies, respectively. The motional frequencies of the particle can be determined accurately through the image currents induced in the trap electrodes 
by the charge's motion~\cite{shockley_currents_1938,wineland_principles_1975}. They are related to the free cyclotron frequency 
\begin{equation}
\label{cyc_freq}
\nu_{c}=\frac{1}{2\pi}\frac{\lvert q \rvert}{m}B,
\end{equation}
which depends on the magnitude $B=|\vec{B}|$ of the magnetic field strength and the charge-to-mass ratio of the particle, through the invariance theorem~\cite{brown_precision_1982} according to
\begin{equation}
\label{inv_the}
\nu_{c}=\sqrt{\nu_+^2+\nu_z^2+\nu_-^2}.
\end{equation}

In contrast to the orbital oscillation frequencies of a trapped particle, the spin-precession frequency for particles with magnetic moments, so-called Larmor frequency $\nu_L$, cannot be determined directly with this image-current detection technique. Instead, the continuous Stern--Gerlach effect is used~\cite{dehmelt_continuous_1986}, where an inhomogeneous magnetic field with $z$ component $B_{z}(z,\rho)=B_{0}+B_{2}(z^2-\rho^2/2)$, a so-called magnetic bottle, is superimposed on the trap center. This magnetic bottle couples the spin magnetic moment to the axial frequency and causes an observable shift of the axial frequency due to a spin transition~\cite{ulmer_observation_2011}. Taking into account the net magnetic moment of the trapped particle, which is composed of its spin magnetic moment and the magnetic moment due to the two radial motions, the resulting shift in the axial frequency is given by 
\begin{multline}
\label{eq:axial_shift}
\Delta\nu_{z,SF}=\frac{\hbar}{2\pi m}\frac{B_{2}}{B_0}\frac{\nu_+}{\nu_{z}}\\
\times \left[ \left(n_{+} + \frac{1}{2}\right) - \frac{\nu_-}{\nu_+} \left(n_- +\frac{1}{2}\right) + m_s\frac{g}{2} \right],
\end{multline}
where $\hbar$ is the reduced Planck constant and $n_{\pm}$ and $m_{s}$ are the quantum numbers associated with the modified cyclotron and magnetron motion and spin orientation, respectively. The Larmor frequency is then determined by measuring spin-flip probabilities for different externally applied drive frequencies through observation of the associated jumps in the axial frequency.

To observe clear jumps of the axial frequency due to a change in $m_s$, spurious noise-driven transitions of $n_+$ (which scale with the value of $n_+$ directly) need to be suppressed so that the resulting frequency changes do not mask the effect of spin flips. Typically, this requires a thermal occupation of the cyclotron motion corresponding to a temperature of about 100\,mK or below~\cite{schneider_double-trap_2017}. Cooling to such low temperatures is realized by thermalizing the cyclotron motion with a tank circuit at liquid-helium temperature for some time, and then monitoring the axial frequency in an alternating fashion. Only when the energy in the modified cyclotron oscillator is low enough, experimentation can proceed. Currently the best achieved median time range to prepare particles at modified cyclotron energies that provide spin state detection fidelities of about 85$\%$ is around $20\,$min~\cite{schneider_double-trap_2017,smorra_observation_2017}. 
This step consumes approximately 90$\%$ of the total measurement time budget, 
and thus represents the key target 
for substantial improvements through quantum logic spectroscopy 
by speeding up the above measurement procedure.
 
In a double-trap $g$-factor measurement the spin state of the particle is first initialized in the magnetic bottle trap. Afterwards, the particle is moved to the homogeneous precision trap where the cyclotron frequency is measured while a magnetic rf drive is applied to flip, once resonant, the particle's spin. In a next step the particle is moved back to the analysis trap and the spin state is analyzed. 
This method is used to scan over spin-flip drive frequencies $\nu_{rf}$
yielding the spin-flip probability as a function of the frequency ratio $\nu_{rf}/\nu_L$ 
from which the Larmor frequency can be extracted. 
The relation
\begin{equation}
\label{g_factor}
\frac{g}{2}=\frac{\nu_L}{\nu_c},
\end{equation} 
valid in the SM,
then permits the determination of the $g$-factor
in the homogeneous magnetic field of the precision trap. 

\section{SME sensitivities in Penning-trap experiments}
\label{SME implications}

For a conventional Lorentz-invariant charged Dirac fermion, such as the SM proton, this $g$-factor corresponds to a coupling that parametrizes the strength of the Pauli interaction~\cite{pauli_relativistic_1941} determined via the effective-Lagrangian contribution $\frac{(g-2)}{m}F^{\mu\nu}\bar{\psi}\sigma_{\mu\nu}\psi$, where $F^{\mu\nu}$ is the electromagnetic field strength and $\psi$ is a single Dirac fermion field. The CPT theorem guarantees that all models of this type are CPT symmetric, so that both a fermion and its associated antifermion necessarily possess the same $g$. But this reasoning also implies that CPT violation requires a framework different from the SM, in which the derivation of Eq.~(\ref{g_factor}) is likely invalidated. Thus, for a clean interpretation and presentation of the proton--antiproton $\nu_L/\nu_c$ comparison, a more meaningful figure of merit might involve coefficients of a suitable test framework that allows for departures from CPT invariance. 

If the test framework is to be a realistic field theory with conventional quantum mechanics, rigorous arguments can be made that CPT violation must be accompanied by deviations from Lorentz symmetry~\cite{greenberg_cpt_2002}. Such a framework can therefore be expected to predict a dependence of $\nu_L/\nu_c$ on the orientation and velocity of the Penning trap. This expectation is borne out by the aforementioned SME, a framework that allows the general classification of such orientation and velocity dependence in all physical systems. The basic idea is that in addition to established physics, the SME Lagrangian contains small Lorentz-violating corrections parametrized by preferred directions in the form of fixed vectors and tensors~\cite{ding_lorentz-violating_2016}:
$b_{p}^{\mu},\, 
H_{p}^{\mu\nu},\,
b_{p}^{(5)\lambda\mu\nu},\,
d_{F,p}^{(6)\kappa\lambda\mu\nu},\,
b_{F,p}^{(5)\lambda\mu\nu},\, 
g_{F,p}^{(6)\kappa\lambda\mu\nu\rho},$ etc.
The components of these vectors and tensors represent the coefficients of the SME, which precision experiments seek to constrain. About half of these coefficients also break CPT symmetry. The notation is chosen such that key physical properties become apparent. For example, the subscript $p$ signifies that the coefficient is associated with the proton. We remark that just like $m$ and $g$, these coefficients cannot be specified independently for a particle and its antiparticle. The subscript $F$ signifies proportionality to $F^{\mu\nu}$, which means in the present context that its physical effects will grow with the trapping $\vec{B}$ field, and the superscript in parentheses indicates mass dimension~\cite{ding_lorentz-violating_2016}. 

With these considerations, one generically expects $\nu_L/\nu_c$ to be given by
\begin{align}\label{GenericPrediction}
2\left(\frac{\nu_L}{\nu_c}\right)_p & =
g+\delta f(b_{p}^{\mu},g_{p}^{\lambda\mu\nu},b_{p}^{(5)\lambda\mu\nu},\ldots)\,,\nonumber\\
2\left(\frac{\nu_L}{\nu_c}\right)_{\bar{p}} & =
g+\delta \bar{f}(b_{p}^{\mu},g_{p}^{\lambda\mu\nu},b_{p}^{(5)\lambda\mu\nu},\ldots)\,.
\end{align}
The CPT-violating corrections are encoded in the functions $\delta f$ and $\delta \bar{f}$, which can in general differ from one another; they can also contain additional physical quantities, such as the proton mass $m$ and the trapping field $\vec{B}$. We note again that both expressions depend on the same set of proton coefficients 
$g,\,m,\,b_{p}^{\mu},\, 
g_{p}^{\lambda\mu\nu},\,
b_{p}^{(5)\lambda\mu\nu}$ etc.
Explicit expressions for $\delta f$ and $\delta \bar{f}$ can be inferred from the results in Ref.~\cite{ding_lorentz-violating_2016}. In the remainder of this section, we present a brief overview of these results.

We begin by considering an inertial Penning trap. The $\vec{B}$ field of such a trap will exhibit a fixed orientation with respect to the SME preferred directions 
$b_{p}^{\mu},\, 
g_{p}^{\lambda\mu\nu},\,
b_{p}^{(5)\lambda\mu\nu},\ldots
$~\cite{ding_lorentz-violating_2016}. 
For this system, we may select coordinates such that $\vec{B}=(0,0,B)$ points entirely along the $x_3$ axis. For leading-order SME effects, it can be argued that the electric field in the Penning trap can be neglected and that dominant SME contributions 
are given by corrections $\delta E_{n,s}^p$ to the usual homogenous-$\vec{B}$ Landau levels~\cite{ding_lorentz-violating_2016}. 

With this setup, one finds for the proton~\cite{ding_lorentz-violating_2016}:
\begin{equation}\label{Cyclotron_F_Correction}
\nu^p_c-\nu^{\bar{p}}_c = 0 + \textrm{higher order.}
\end{equation}
Although each stack of Landau levels, determined by the spin state, is affected differently, the dominant SME shifts {\it within} a fixed stack are identical for each level, so that the transition frequencies $\nu^p_c$ and $\nu^{\bar{p}}_c$ are unaffected within the SME at leading order. For subdominant effects, see Ref.~\cite{ding_lorentz_2020}.
 
On the other hand, spin-flip transitions {\it between} Landau stacks do contain leading-order corrections:
\begin{align}\label{Larmor_F_Correction}
\nu^p_L =
&\, 
+2\tilde{b}^{3}_p-2\tilde{b}^{33}_{F,p}B\,,
\nonumber\\
\nu^{\bar{p}}_L =
&\, 
-2\tilde{b}^{*3'}_p+2\tilde{b}^{*3'3'}_{F,p}B^*\,.
\end{align}
Here, the $x_{3'}$ axis denotes the direction of the $\vec{B}^*$ field in the antiproton Penning trap, which may differ in magnitude and direction from $\vec{B}$ in the proton trap. The tilde coefficients represent combinations of preferred-direction coefficients in the SME: 
\begin{align}\label{SMEcoeffs}
\tilde{b}_p^{3} =
&\;
b_p^{3}+H_p^{12}
- m\big[d_p^{30}+g_p^{120}\big]\nonumber\\
&
{}+ m^2\big[b_p^{(5)300}+H_p^{(5)1200}\big]
\nonumber\\
&
{}-m^3\big[d_p^{(6)3000}+g_p^{(6)12000}\big],
\nonumber\\
\tilde{b}_{F,p}^{33} =
&\;
b_{F,p}^{(5)312}+H_{F,p}^{(5)1212}
- m\big[d_{F,p}^{(6)3012}+g_{F,p}^{(6)12012}\big],
\end{align}
and 
\begin{align}\label{SMETildecoeffs}
\tilde{b}_p^{*3} =
&\;
b_p^{3}-H_p^{12}
+ m\big[d_p^{30}-g_p^{120}\big]\nonumber\\
&
{}+ m^2\big[b_p^{(5)300}-H_p^{(5)1200}\big]
\nonumber\\
&
{}+m^3\big[d_p^{(6)3000}-g_p^{(6)12000}\big],
\nonumber\\
\tilde{b}_{F,p}^{*33} =
&\;
b_{F,p}^{(5)312}-H_{F,p}^{(5)1212}
+ m\big[d_{F,p}^{(6)3012}-g_{F,p}^{(6)12012}\big].
\end{align}
The grouping of SME coefficients in these expression has been chosen so as to display their properties under rotations, as will become apparent below.

Since Lorentz- and CPT-violating directions represent 4-vectors and 4-tensors, their components can in principle be specified with respect to any basis. However, for practical reasons, such as for comparison between different experiments, measurements are quoted with respect to a standard coordinate system. The canonical choice for this system is the Sun-centered celestial equatorial frame (SCCEF)~\cite{kostelecky_signals_2002}. The above tilde components 
$\tilde{b}^{3}_p$,
$\tilde{b}^{33}_{F,p}$,
$\tilde{b}^{*3}_p$, and
$\tilde{b}^{*33}_{F,p}$
must then be expressed in terms of SCCEF components, conventionally denoted with capital coordinates and indices $X^J=(X,Y,Z)$. For usual terrestrial Penning traps, the Lorentz transformation between the laboratory and the SCCEF can be decomposed into various contributions including the rotations and boosts arising from Earth's revolution about its axis and its orbital motion around the Sun. Below, we focus on the primary contribution only: the rotation transformation due to the laboratory's motion about the Earth's axis. Purely non-inertial effects due to the rotation of the laboratory frame, such as Coriolis-type corrections, are small and will be neglected. In this context, it is useful to introduce also standard local laboratory frames with a notation involving lower-case coordinates and indices $x^j=(x,y,z)$ with the corresponding unit vectors $(\hat{x},\hat{y},\hat{z})$; the conventional choice of these axes has $\hat{z}$ pointing straight up to the local zenith, and $\hat{x}$ and $\hat{y}$ are oriented horizontally and point in the local southern and eastern directions, respectively. In these coordinates, $\vec{B}$ may no longer point along the $z$ direction, but rather along $\hat{n}=\vec{B}/B=(n^1,n^2,n^3)$. For example, $\hat{n}=(0,0,1)$ for $\vec{B}$ vertically upward, or $\hat{n}=-(\cos\xi,\sin\xi,0)$ for horizontal $\vec{B}$ at an angle $\xi$ west of north. The conversion between such laboratory frames and SCCEF coordinates is then given by
\begin{align}\label{rot}
R^{jJ}(\chi,T_{\oplus})=
&\nonumber\\
&\hspace{-14mm}
\left(\begin{array}{ccc}
\cos\chi \cos\omega_{\oplus} T_{\oplus} 
& 
\cos\chi \sin\omega_{\oplus} T_{\oplus} 
& 
-\sin\chi
\\
-\sin\omega_{\oplus} T_{\oplus} 
& 
\cos\omega_{\oplus} T_{\oplus} 
& 
0 
\\
\sin\chi \cos\omega_{\oplus} T_{\oplus} 
& 
\sin\chi \sin\omega_{\oplus} T_{\oplus} 
& 
\cos\chi 
\end{array} \right).
\end{align}
Here, $\chi$ denotes the colatitude of the laboratory, $\omega_{\oplus}=2\pi/(23\,\textrm{h}\;56\,\textrm{min})$ the Earth's sidereal angular frequency, and $T_{\oplus}$ the local sidereal time~\cite{ding_lorentz-violating_2016}. 

With the above definitions, the Penning-trap frequency shifts~(\ref{Larmor_F_Correction}) can be expressed with SCCEF components as follows:
\begin{align}\label{SMETildecoeffs}
\tilde{b}_p^{3} =
&\;
n^{j} R^{jJ}(\chi,T_{\oplus})\;\tilde{b}_p^{J},
\nonumber\\
\tilde{b}_{F,p}^{33} =
&\;
n^{j} R^{jJ}(\chi,T_{\oplus})\;
\tilde{b}_{F,p}^{JK}\,
R^{Kl}(\chi,T_{\oplus})\;n^{l},
\nonumber\\
\tilde{b}_p^{*3} =
&\;
n^{j'} R^{j'J}(\chi^*,T^*_{\oplus})\;\tilde{b}_p^{*J},
\nonumber\\
\tilde{b}_{F,p}^{*33} =
&\;
n^{j'} R^{j'J}(\chi^*,T^*_{\oplus})\;
\tilde{b}_{F,p}^{*JK}\,
R^{Kl'}(\chi^*,T^*_{\oplus})\;n^{l'}\,,
\end{align}
where starred or primed quantities refer to the antiproton laboratory, which may differ from the proton laboratory. With these results, explicit expressions for the corrections $\delta f$ and $\delta\bar{f}$ in Eq.~(\ref{GenericPrediction}) can be obtained. Note that  these expressions imply an oscillatory time dependence of the observable $2\nu_L/\nu_c$  
in the form of sidereal variations~\cite{ding_lorentz-violating_2016}. Note also the dependence on the laboratory's location via both colatitude and local sidereal time, as well as the dependence on the $\vec{B}$-field direction and magnitude.

\section{Experimental concept for quantum logic spectroscopy}
\label{Experimental concept}

\begin{figure*}[t]
	\centering
	\includegraphics[width=0.98\textwidth]{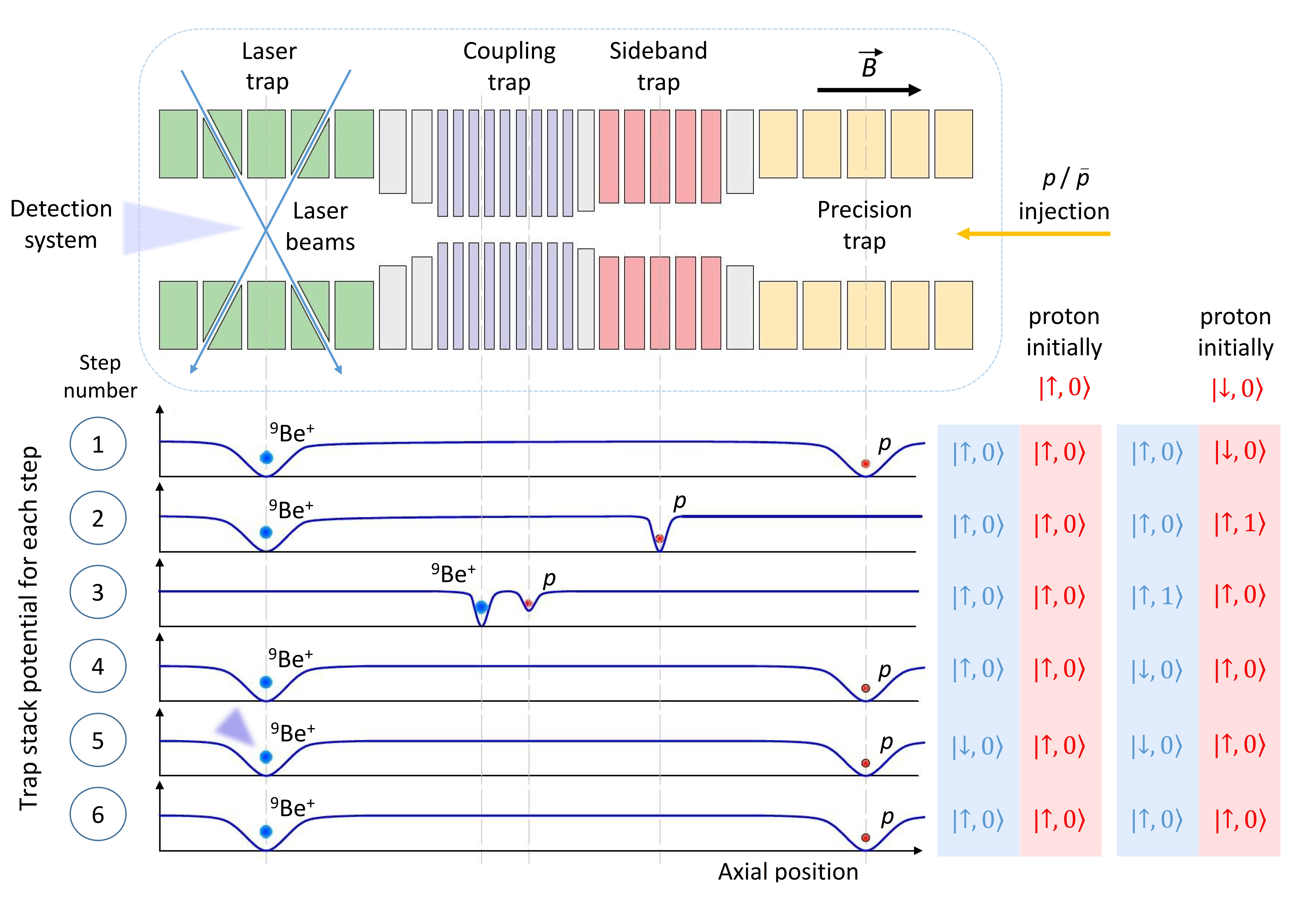}
	\caption{Penning trap system and experimental protocol. Top: Different traps are used for the implementation of the technique. The laser trap for beryllium ions is represented in green. The coupling trap to produce the double-well potential is depicted in purple. The red and light orange electrodes are used to couple their spin and motional states and to confine single protons or antiprotons, respectively. Finally, the gray electrodes are utilized for adiabatic transport between the traps. Bottom: Sketch of the electric potential along the magnetic field axis during the different steps of the experimental protocol. Spin and motional states at the end of steps 1--6 are depicted for an initial $\mid \uparrow \rangle$ and $\mid \downarrow \rangle$ spin state of the proton on the left and right, respectively. Further details about the experimental procedure are given in the main text. Drawing is not to scale.}
	\label{fig:trap_system}
\end{figure*}

From the discussion in Sec.~\ref{Present status}, 
it should be evident that $g$-factor measurements with \ppbars 
are extremely challenging because of the extraordinarily low noise levels required, the long cooling times, and the need to have outstanding control of systematic effects in the presence of considerable residual magnetic-field inhomogeneities. Measurements of sidereal variations in proton--antiproton $\nu_L/\nu_c$ comparisons introduce an additional layer of complexity 
because of the amount of data required and the complication due to the interleaved long cooling times. Different sympathetic laser-cooling approaches have already the potential to reduce considerably the time required for cooling. Here, in addition, we discuss prerequisites and approaches for the implementation of quantum logic spectroscopy, which would allow one to probe all resonances essentially from the motional ground state of the particle, with cooling or spin-state detection happening on time scales of a second or less. The underlying principles were suggested in Refs.~\cite{heinzen_quantum-limited_1990,wineland_experimental_1998-1} and result in the following protocol,
which is also illustrated in the bottom right part of Fig.~\ref{fig:trap_system}.

\begin{enumerate}
	\item Both particles are assumed to be cooled to the ground state and the `logic' ion (blue) is prepared in the $\mid \uparrow \rangle$ state as shown in Fig.~\ref{fig:trap_system}. The spin state of the \ppbar (red) is unknown.
	\item An rf pulse is applied to flip the \ppbar spin if and only if the \ppbar spin-state was \mbox{$\mid\downarrow\rangle$} previously, and to add one quantum of motion in the axial mode. This effectively implements a SWAP gate between the \ppbarss spin and motional ($n_p=0$ vs.\ $n_p=1$) qubit ~\cite{nitzschke_elementary_2020}. 
	\item The motional state of the \ppbar is transferred to the `logic' ion by a double-well potential technique~\cite{wineland_experimental_1998-1,niemann_cryogenic_2019}. This realizes a SWAP gate between the \ppbarss and the `logic' ion's motional ($n_l=0$ vs.\ $n_l=1$) qubits. 
	\item A laser pulse flips the `logic' ion's spin state and removes a quantum of motion if and only if it was previously in the $n_l=1$ state. This realizes a SWAP gate between the `logic' ion's motional qubit and its spin state. 
	\item The fluorescence signal induced by a resonant laser beam is used to detect the spin state of the `logic' ion. If and only if the ion is $\mid \uparrow \rangle$ at the end of the previous step, scattered photons are detected.  
	\item The `logic' ion is re-initialized to $\mid \uparrow \rangle$ by optical pumping, and the motion of both particles is re-initialized to the ground state, if necessary. 
\end{enumerate} 
Repetition of steps 1 to 6, accompanied by variation of the frequency of a rf pulse to flip the \ppbar spin in step 1 yields a resonance curve from which $\nu_L$ can be extracted. 

Implementing these ideas requires several ingredients:
\begin{enumerate}[a.]
  \item Spin-state discrimination for the `logic' ion
	\item The ability to ground-state cool the `logic' ion and carry out spin--motional state swap operations on the `logic' ion
	\item A suitable trap stack for proton loading and antiproton injection
	\item A suitable coupling mechanism allowing the transfer of single quanta of motion between the `logic' ion and the \ppbar
	\item The ability to carry out spin--motional state swap operations on \ppbars
	\item Means of probing the motional modes and irradiating radiation to drive the spin-flip transition for high-precision spectroscopy
	\item Motional ground-state transport for both \ppbars and the `logic' ion if not all operations are carried out in a single trap zone
\end{enumerate}

Based on these requirements, a multi-trap method is favoured. Different points above impose mutually incompatible boundary conditions difficult to meet in a single trap. For example, probing the Larmor or cyclotron resonance would require a trap of relatively large scale (typically 5--10\,mm), such that the systematics are well controlled~\cite{schneider_double-trap_2017}. Laser access, on the other hand, typically involves modifications to the trap geometry which slightly distort potentials and impose additional systematic effects~\cite{brown_geonium_1986}. Furthermore, laser-induced charging may inhibit the refined control of potentials necessary for a precision trap~\cite{harlander_trapped-ion_2010}. Implementing a coupling mechanism between single \ppbars and the `logic' ion at the single motional quantum level that works in a reasonably short amount of time typically requires a small length scale for the trap~\cite{smorra_base_2015}, so that both particles can be kept in close proximity, which is again at odds with the demands of a precision trap. Implementing the swap operation between the \ppbarss spin and motional states would typically also call for a small-scale trap and additional coupling fields~\cite{wineland_experimental_1998-1}, which might again conflict with the needs for a precision trap. 

Looking at the list of requirements, the core functions can be realized in a trap stack, schematically illustrated in the upper left of Fig.~\ref{fig:trap_system}, with four different zones for these functions. Items (a) and (b) can be carried out in a trap zone with laser-beam access and fluorescence detection. To allow for item (c), antiproton loading, one end of the trap stack needs to be open for the injection of antiprotons. Item (d) could be accomplished in a dedicated coupling trap with small length scales in order to reduce the distance between the `logic' ion and the \ppbar. Item (e) could be performed in a sideband trap for spin-motional coupling with \ppbars, inspired by recent advances in rf or microwave driven quantum logic operations for trapped-ion qubits~\cite{nitzschke_elementary_2020}. Item (f) could be realized using a precision trap as employed in present experiments. Below we discuss these different trap zones before highlighting how they come together to allow the implementation of quantum logic spectroscopy. These traps would be interconnected through physical transport of particles along the magnetic-field axis (item (g) above). In order to facilitate transport, traps should be built as cylindrical Penning traps. The physical location of both particles during each step of the measurement algorithm is indicated in the lower left part of Fig.~\ref{fig:trap_system}. 

We start by discussing the general structure of the apparatus. As in present $g$-factor measurements~\cite{smorra_parts-per-billion_2017, schneider_double-trap_2017}, the experiment should be operated at cryogenic temperatures to facilitate pre-cooling of \ppbars using cryogenic tank circuits, and it should feature a hermetically sealed secondary vacuum enclosure (`trap can') to support long storage times of antiprotons. In order to obtain magnetic fields as homogeneous as possible, radial optical access ports in the Penning-trap's superconducting magnet should be avoided, as they are likely to introduce additional inhomogeneities in the magnetic field~\cite{gruber_formation_2005}. As a consequence, all laser beams would have to be brought in along the axis of the superconducting magnet, and fluorescence detection would have to occur along that direction as well, and thus downstream from the antiproton source. The laser cooling and detection trap would thereby also have to be the last trap in the trap stack with the imaging optics downstream of it. 
In what follows, we consider a magnetic field of $5\,$T for higher cyclotron and Larmor frequencies and beryllium ions are assumed as 'logic' ions of choice.

\subsubsection{Laser cooling and detection trap}
 
\begin{figure}[t]
	\centering
	\includegraphics[width=0.5\textwidth]{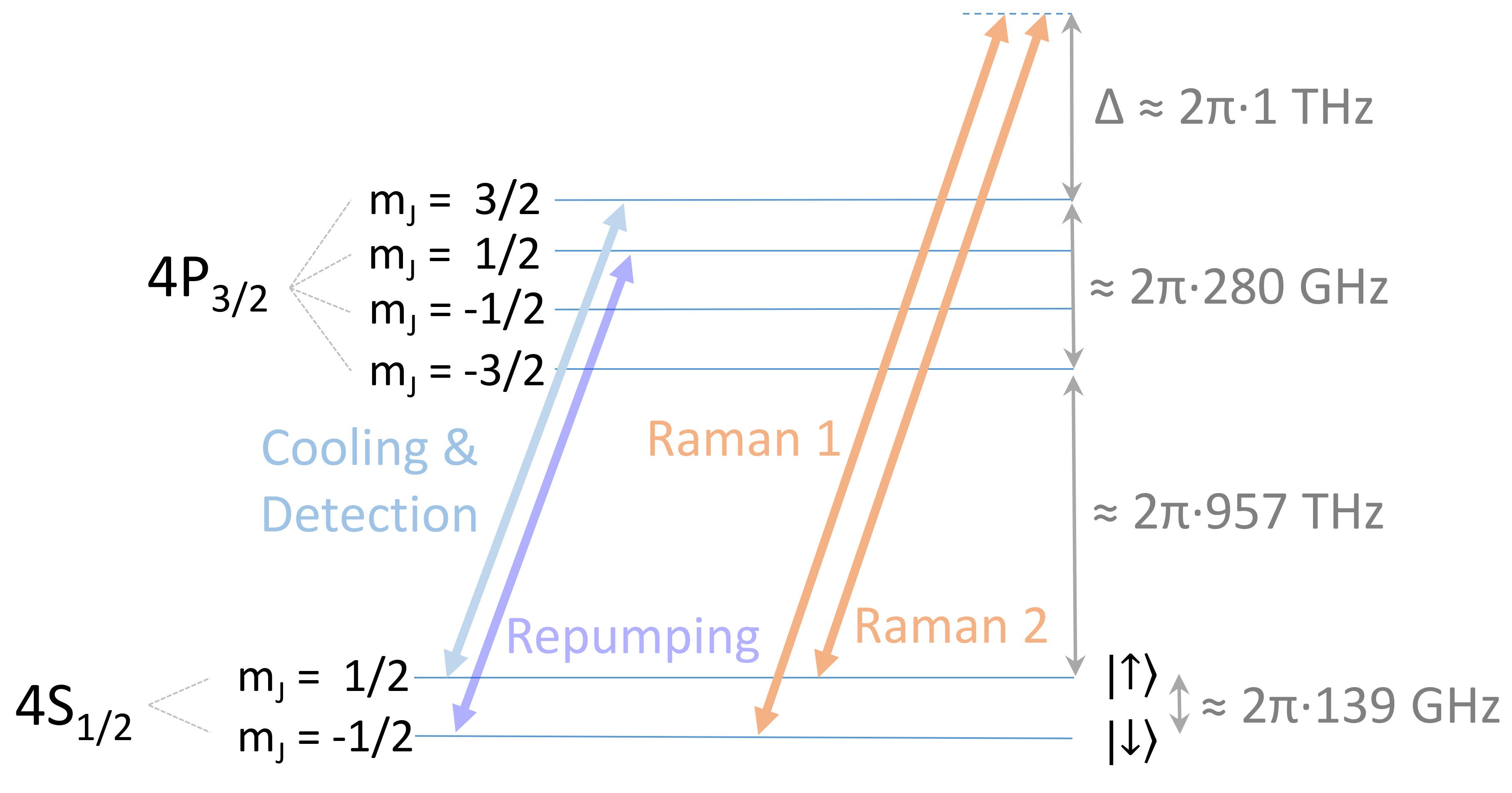}
	\caption{Ilustration of the energy level of \Be at a 5~T magnetic field. The transitions used for Doppler cooling and repumping as well as for the two-photon stimulated Raman process are shown. Laser detunings should be chosen for the $m_I=+3/2$ nuclear spin state. Weak off-resonant transitions will eventually lead to the preparation of that nuclear spin state starting from $m_I\ne+3/2$. Energy levels are not drawn to scale.}
	\label{fig:lasers}
\end{figure}

The design of our current laser cooling and detection trap has been discussed in Ref.~\cite{niemann_cryogenic_2019}. Beryllium ions are produced in the center of the laser trap by ablation of beryllium atoms from a solid target. A single trapped ion can be obtained from repeated splitting of the loaded sample, using suitable voltage ramps, and then discarding one part. This ion can be laser-cooled near the Doppler limit using two laser fields resonant with the $\ket{^2S_{1/2}, m_J=+1/2}$ $\rightarrow$ $\ket{^2P_{3/2}, m_J=+3/2}$ and $\ket{^2S_{1/2}, m_J=-1/2}$ $\rightarrow$ $\ket{^2P_{3/2}, m_J=+1/2}$ transitions in \Be as is shown in Fig.~\ref{fig:lasers}. Details of the laser-system setup for this experimental arrangement can be found in Refs.~\cite{meiners_towards_2017-2,niemann_cryogenic_2019}. 

\subsubsection{Ground-state cooling and spin-motional SWAP gate for `logic' ion}
\label{Ground-state cooling}
Ground-state cooling and the implementation of the SWAP gate between the ion's spin and motional qubits will require the implementation of red and blue spin-motional sideband pulses ~\cite{diedrich_laser_1989}. In the case of \Be, these will have to be carried out on the Zeeman qubit in the \Be ground state, split by roughly $139\,\mathrm{GHz}$ at $5\,$T. In Ref.~\cite{paschke_versatile_2019}, we have demonstrated a pulsed-laser approach to bridge this big Zeeman splitting by using two laser beams as depicted in Fig.~\ref{fig:lasers}, and successfully implemented it at a low magnetic field of 22.3~mT. For sideband cooling, blue-sideband pulses on the $\ket{^2S_{1/2}, m_J=+1/2}$ to $\ket{^2S_{1/2}, m_J=-1/2}$ transition can be employed to remove motional energy from the ion, followed by repump pulses on the $\ket{^2S_{1/2}, m_J=-1/2}$ to $\ket{^2P_{3/2}, m_J=+1/2}$ transition to re-initialize the ion in the $\ket{^2S_{1/2}, m_J=+1/2}$ state. In order for the laser beams to couple only to the axial direction for sideband cooling, they need to be irradiated with a wave-vector difference along the trap axis as illustrated in Fig.~\ref{fig:trap_system}.

\subsubsection{Proton loading and antiproton injection}
With one end of the trap stack open, antiprotons can be injected and trapped using a catching and reservoir trap as in the BASE-CERN setup~\cite{smorra_base_2015}. Proton loading will have to be carried out differently. In the BASE-CERN setup, the electron gun used for this operation is pointing in the opposite direction as the antiproton beam, a direction obstructed by the fluorescence detection system in our setup. We therefore plan to use an off-axis electron gun and an organic target for proton loading~\cite{pulido_proton_2019-1}. \Ppbars could be stored for use in a reservoir trap similar to that demonstrated at BASE-CERN~\cite{smorra_reservoir_2015}.

\begin{figure*}[t]
	\centering
	\includegraphics[width=1.4\columnwidth]{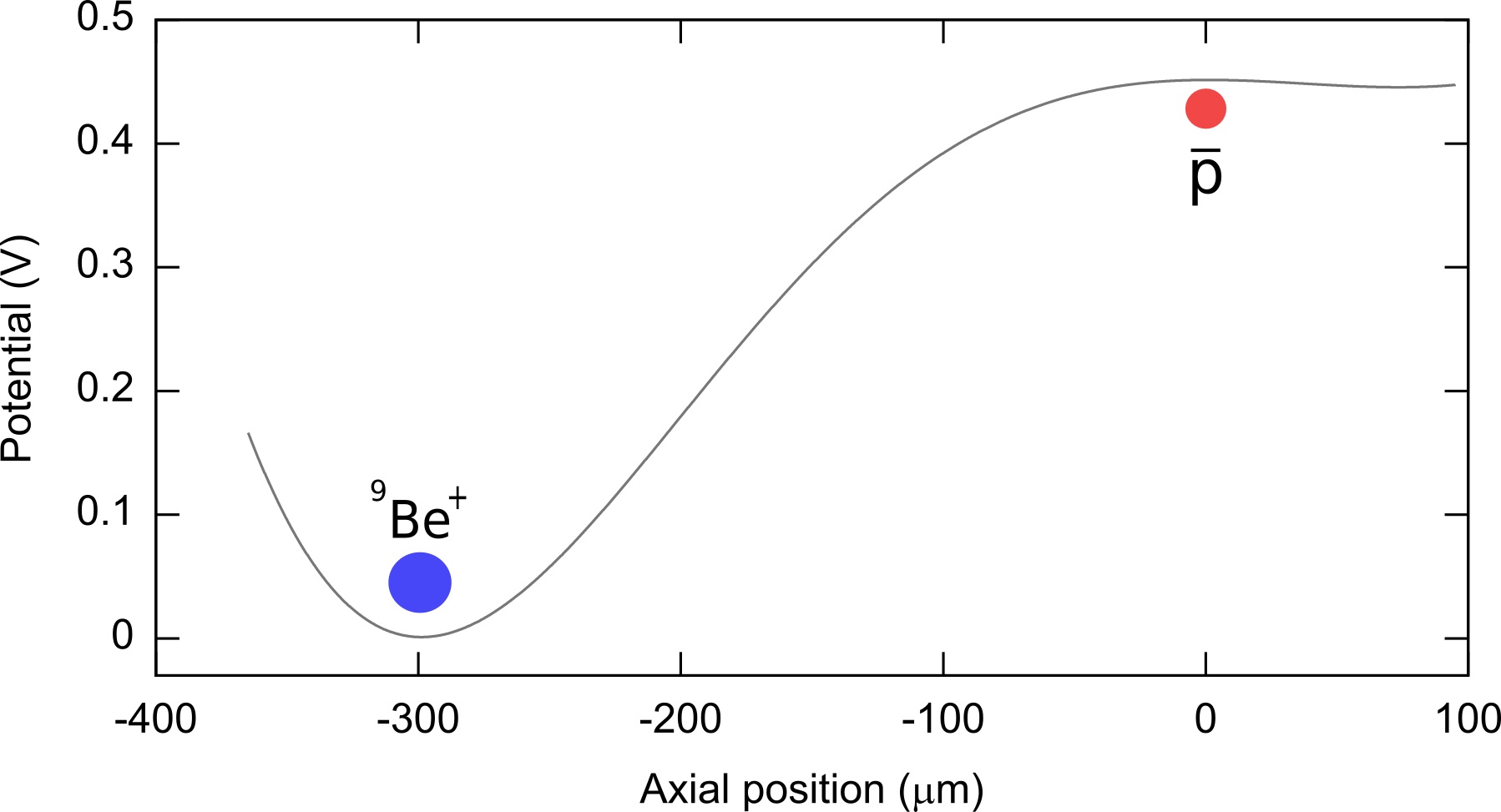}
	\caption{Potential shape along the $z$ axis for storing a single \Be ion (blue) and a single antiproton (red)  at a distance of $300\,\mu$m. The coupling trap used to simulate this potential shape is composed of nine ring-shaped electrodes with an inner diameter of $800\,\mu$m and a thickness of $200\,\mu$m each. The distance between adjacent electrodes is $50\,\mu$m. The potential shape has been calculated for an axial frequency of $4\,$MHz for both particles in a $5\,$T Penning trap. The potential shape for the coupling between a \Be ion 	and a single proton was reported in Ref.~\cite{smorra_base_2015}.}
	\label{fig:microtrap}
\end{figure*}

\subsubsection{Coulomb coupling using a double-well potential}
\label{Double-well potential}
A crucial feature of our experimental protocol is the coupling between a single \ppbar and a laser-cooled \Be ion through the Coulomb interaction in free space. For this purpose, the two particles should be spatially separated and held in adjacent potential wells. We have already discussed this ion coupling mechanism in Ref.~\cite{meiners_towards_2018, niemann_cryogenic_2019, smorra_base_2015}. For two charges $q_1$ and $q_2$ trapped in two potential wells separated by a distance $s_0$, if the radial component of the Coulomb force is neglected and a one-dimensional interaction is assumed, in the limit of weak coupling with both particles oscillating at the same axial frequency $\nu_z \simeq \nu_1 = \nu_2$, the particles exchange energy at the time scale
\begin{equation}
\label{exchange_time}
t_{\rm ex}=\frac{\pi}{2}\frac{1}{\Omega_{\rm ex}}=\frac{2\pi^2\epsilon_0s_0^3\nu_z\sqrt{m_1m_2}}{q_1q_2}\,,
\end{equation} 
where $\Omega_{\rm ex}$ is the so-called coupling rate, $\epsilon_0$ is the permittivity of free space, and $m_1$ and $m_2$ are the masses of particle 1 and 2, respectively. This coupling mechanism has been demonstrated for two \Be ions in their motional ground states in a surface-electrode rf trap~\cite{brown_coupled_2011} and for \Ca ions in Ref.~\cite{harlander_trapped-ion_2011}; it should also be feasible in Penning traps because both systems are analogous in the axial mode. Figure~\ref{fig:microtrap} shows the double-well potential shape along the axial direction for storing a \Be ion and a single antiproton. For an axial frequency $\nu_z = 4\,$MHz, a distance $s_0 = 300\,\mu$m leads to an exchange time of $t_{\rm ex} \simeq 3.7\,$ms~\cite{smorra_base_2015}.

The choice of \Be as the `logic' ion implies higher coupling rates due to its low mass compared to other readily laser-cooled ions. Also, it is desirable to have the ratio of the `logic' ion mass to the proton mass as close to unity as possible in order to facilitate the engineering of a double-well potential with two equal trap frequencies for both particles, which requires a potential-curvature ratio given by the mass ratio. Another advantage of \Be is the relatively simple atomic structure without metastable states which would otherwise require a substantial amount of repumping fields as in the case of \Ca. On the other hand, for ground-state cooling, \Be requires stimulated Raman transitions, which can be challenging to implement given the Zeeman splittings in the Penning trap that typically exceed $100\,$GHz. 

A further essential feature for achieving high coupling rates is a small distance $s_0$ between the two trapped particles, as is apparent from Eq.~(\ref{exchange_time}). The limitations in this respect are the trap construction together with anomalous motional heating rates~\cite{turchette_heating_2000}, and the initial amplitude of radial motion of the \ppbars before they are brought into the coupling trap. To place the minima of the potential wells closer together in a cylindrical Penning trap, both thickness and diameter of the electrodes must be reduced. Microfabrication techniques can be employed for the construction of such traps~\cite{cornejo_optimized_2016}, allowing distances $s_0$ on the order of a few hundred $\mu$m. However, this involves reducing the trap diameter, which would lead to an increase of the anomalous motional heating rate. 

The cryogenic operation of the trap will suppress the anomalous motional heating. Recently, rates of a few quanta per hour have been reported for the motional heating of a single antiproton in a cryogenic Penning trap with a $3.6\,$mm inner diameter~\cite{borchert_measurement_2019}. Assuming that magnetron amplitudes of \ppbars injected into the coupling trap could be on the order of a few tens of microns, it would seem reasonable to consider an ion-to-electrode distance in the coupling trap of no more than 400--500\,$\mu$m.

Using such a double-well trap, sympathetic cooling of single \ppbars to the motional ground state could proceed as follows:
\begin{enumerate}[1.]
	\item The \Be ion is cooled to the motional ground state by using Doppler~\cite{wineland_radiation-pressure_1978} and sideband cooling~\cite{goodwin_resolved-sideband_2016} techniques as discussed above. A single \ppbar is prepared in the precision trap and pre-cooled by a resistive cooling technique.   
	\item Both particles are adiabatically shuttled to the coupling trap. The axial motional modes of both particles are tuned into resonance for the time it takes to swap the motional states between the two particles.
	\item The `logic' ion is transported back to the laser cooling trap and re-cooled to the motional ground state using the procedure outlined in step 1. 
\end{enumerate} 
This implements sympathetic cooling of the \ppbar to the motional ground state. The relevant time scales would be the time required for motional ground state cooling of the \Be ions (a few milliseconds at most)  and the motional exchange time of $t_\mathrm{ex}=3.7\,\mathrm{ms}$, enabling ground state cooling of the axial degree of freedom of the \ppbar within milliseconds, significantly shorter than the time scales associated with resistive cooling. 

The motional state exchange could exhibit imperfections if the initially finite magnetron radius of the \ppbar (see above) leads to a different effective distance of the two particles. This effect should be small, because an \ppbar magnetron radius of a few tens of microns would not effectively change the relevant distance of the two particles of e.g.\ $300\,\mu\mathrm{m}$ by much. Other imperfections could arise from timing and resonance-condition imperfections or due to trap anharmonicities. 
\begin{enumerate}[1.]
	\setcounter{enumi}{3}
	\item In the case of imperfect cooling, steps 2 and 3 could be performed several times. 
	\item To cool the radial modes, a sideband coupling procedure between the radial and axial modes of the \ppbar must be carried out in the precision trap~\cite{cornell_mode_1990}, followed by repetition of the above axial-cooling protocol.
\end{enumerate} 

\subsubsection{Sideband coupling between motional and spin states}
\label{Sideband coupling}
To couple the motional and spin states of the \ppbar in the sideband trap and to carry out the SWAP operation between the \ppbarss spin state and its motional qubit, a rf pulse close to the \ppbar spin flip resonance $\omega_{SF} = g_p \mu_N B / \hbar$ should be applied, where $\mu_N$ is the nuclear magneton and $g_p$ is the \ppbar gyromagnetic ratio in units of $\mu_N$. For this purpose, the proposal of Ref.~\cite{heinzen_quantum-limited_1990} envisioned the use of an oscillating-amplitude magnetic-field gradient to couple spin and motional degrees of freedom. This mechanism is conceptually similar to the oscillating near-field microwave quantum logic gates of Ref.~\cite{ospelkaus_microwave_2011}. However, with the particle at least a factor of ten further away from the conductor inducing the transition, without any current scaling, the gradient would be about two orders of magnitude smaller. In addition, compared to an electronic atomic magnetic moment, the magnetic moment of the \ppbar is lower by about three orders of magnitude, making effective Rabi rates about five orders of magnitude smaller. 

This issue can be overcome by introducing a strong static magnetic-field gradient~\cite{mintert_ion-trap_2001}. In a variation of the magnetic bottle approach, very large magnetic field gradients can be generated in the center of a Penning trap by introducing a ferromagnetic electrode as an endcap electrode, effectively trapping the particle in the wing of a magnetic bottle, where the magnetic-field curvature vanishes and the gradient is strongest. For traps with an inner diameter of about 1\,mm, gradients as strong as 1000\,T/m can be anticipated, much stronger than the $\approx 35\,\mathrm{T/m}$ that was used in Ref.~\cite{ospelkaus_microwave_2011}, yet at considerably larger surface-to-electrode distance, thereby at least partly compensating for the smallness of the magnetic moment of the \ppbar. In Ref.~\cite{nitzschke_elementary_2020}, we have analyzed the behavior of an \ppbar in a large magnetic-field gradient superimposed on the Penning trap and found that as in Ref.~\cite{mintert_ion-trap_2001}, a homogeneous magnetic rf field at a spin-motional sideband frequency will now allow the spin and motional degrees of freedom to be coupled as required for the implementation of the SWAP gate. 

\subsubsection{Cyclotron-frequency measurement}
In addition to the Larmor frequency, a $g$-factor determination based on Eq.~\ref{g_factor} also requires a cyclotron-frequency measurement for the \ppbar. For this purpose, different approaches may be taken. One of these is based on the image-current technique in the precision trap. However, this method requires the particle to be in thermal equilibrium with the detection system~\cite{wineland_principles_1975}, which may be a disadvantage for tests with the highest precision when residual magnetic-field inhomogeneities are present. 

An alternative scheme that probes all transition out of the motional ground state at the level of a single quantum has already been discussed in Refs.~\cite{heinzen_quantum-limited_1990,wineland_experimental_1998-1}. It relies on the application of the same sympathetic ground-state cooling protocol outlined above and on an electric-field pulse to probe the motional resonances in the precision trap. A possible excitation as a result of the pulse can then be read out using the Coulomb-coupling technique and the `logic' ion again. By varying the excitation-pulse frequency, the motional resonances can be recorded and the free cyclotron frequency determined. 

A variant of this method would be carried out by using a Ramsey-like experiment~\cite{ramsey_molecular_1950}, where two electric-field pulses, separated by a free-precession time $t_{f}$, are applied~\cite{mccormick_quantum-enhanced_2019, wolf_motional_2019}. The first excitation pulse produces a coherent state by imposing a displacement $\hat{D}(\alpha) = \exp(\alpha \hat{a}^{\dagger} - \alpha^{*} \hat{a})$ on the motional mode of interest, where $\alpha$ is the displacement amplitude and $\hat{a}$ and $\hat{a}^{\dagger}$ represent the usual creation and annihilation operators of the trap harmonic oscillator. The expected state overlap is given by $\Braket{n|\hat{D}|n} = \rm exp(-1/2|\alpha|^2) \mathcal{L}_n (|\alpha|^2)$, where $\mathcal{L}_n$ is the Laguerre polynomial~\cite{de_oliveira_properties_1990}. After a free period $t_{f}$, a relative phase $\phi = \Delta \times t_{f}$ is accumulated in phase space, where $\Delta$ is the detuning of the excitation. For $\phi=\pi(k+1)$ with $k$ an integer, the second pulse undoes the displacement and returns the system to $\ket{0}$~\cite{wolf_motional_2019}. For any other outcome, the final state will have an excited state contribution different from $\ket{0}$ which can be detected using the axial Coulomb coupling technique, combined with $\pi$-pulses which can exchange the axial and cyclotron degrees of freedom~\cite{cornell_mode_1990}.

\subsubsection{Ion transport}
Regarding ion transport, adiabatic transport is achieved for a harmonic trapping potential with a minimum that can be shuttled along the axial direction between two traps without any abrupt changes to the potentials. In the context of quantum-information processing with linear rf traps, this goal has already been achieved for particles in their motional ground state~\cite{rowe_transport_2002}, even through complex trap geometries~\cite{blakestad_high-fidelity_2009,blakestad_near-ground-state_2011}. In principle, the axial degree of freedom in a Penning trap is purely electrostatic, as for the linear rf trap, so one would expect comparable results. With the methods available in the BASE experiment at CERN, transport heating has been characterized to 0(2) quanta per transport in the cyclotron mode~\cite{stefan_slide_2021}. Stimulated-Raman laser control would be able to place tighter bounds here, and would be expected to yield results similar to Paul traps with significantly less than one quantum of motion per round-trip, as observed in Paul traps. Note that this is a prerequisite to carry out the above protocols. 

\section{Sensitivities and SME implication}
\label{Sensitivities and SME implications}

By using the experimental protocol discussed in this work, a Larmor-frequency determination will consist of several experimental cycles. For each $i$-th cycle, the spin-flip probability is determined for a pulse excitation $\nu_L^{i}$ close to $\nu_L$. The principal steps for a single measurement cycle are summarized in Table~\ref{table:1}. The Larmor resonance is probed by using the Ramsey method with two $\pi/2$ pulses separated by a free-precession time $t_{f}$. Each pulse must be a transverse magnetic rf field at an amplitude $b_{\mathrm{rf}}$ and duration $t_{p} = \pi m/g_pqb_{\mathrm{rf}}\approx 3\,$ms for $b_{rf} \approx 2\,\mu$T~\cite{nagahama_high-precision_2016}, which must be shorter than $t_{f}$. Transport times will depend on the low-pass RC filter of the DC lines, whose cutoff frequency will be given by the magnetron frequency. After that, sideband coupling between the motional and spin states of the \ppbar can be produced in around $2\,$ms with low intrinsic error~\cite{nitzschke_elementary_2020}. Coulomb coupling is discussed in detail in Sec.~\ref{Double-well potential}, and finally, a SWAP gate between the Be ion's motional and internal qubits can be performed in several hundred $\mu$s~\cite{paschke_versatile_2019}, as can photon detection~\cite{brown_coupled_2011}. For $t_{f} \approx 100\,$ms, the probe time is around $90\,\%$ of the total measurement time.       

\begin{table}[t]
	\caption{Time consumption for the main steps of a measurement cycle. The trap where each step will take place is indicated. The beryllium, coupling, sideband, and precision traps are abbreviated as BT, CT, ST, and PT, respectively. See text for further details.}
	\centering
	\begin{tabular}{ l l c r }
		\hline
		\hline
		\multicolumn{4}{c}{Larmor frequency $\nu_L^{i}$} \\ 
		\hline
		\hline
		Step & & Trap  & Time \\
		\hline
		I) & Pulse excitation at $\nu_L^{i}$ &  PT  & 3 ms \\
		II) & Free-precession time &  PT  & 100 ms \\  
		III) & Pulse excitation at $\nu_L^{i}$ &  PT  & 3 ms \\   
		IV) & Proton Transport &  -  & 1 ms \\
		V) & Proton SWAP operation &  ST  & 2 ms \\
		VI) & Proton and Be Transport & -  & 1 ms\\
		VII) & Coulomb coupling & CT  & 4 ms\\
		VIII) & Proton and Be Transport & - & 1 ms\\
		IX) & Be detection & BT & 1 ms\\
		\hline
		\multicolumn{3}{l}{Total time for $\nu_L^{i}$} & $\approx$~116 ms\\
		\hline
		&  &  &\\
		\hline
		\hline
		\multicolumn{4}{c}{Cyclotron frequency $\nu_c$} \\
		\hline
		\hline
		\multicolumn{4}{c}{Axial frequency $\nu_z^{i}$} \\ 
		\hline
		\hline
		Step & & Trap  & Time \\
		\hline
		I) & Pulse excitation at $\nu_z^{i}$ &  PT  & < 1 ms \\
		II) & Free-precession time &  PT  & 100 ms \\  
		III) & Pulse excitation at $\nu_z^{i}$ &  PT  & < 1 ms \\  
		IV) & Proton and Be Transport &  -  & 1 ms \\
		V) & Coulomb coupling & CT  & 4 ms\\
		VI) & Proton and Be Transport & - & 1 ms\\
		VII) & Be detection & BT & 1 ms\\
		\hline
		\multicolumn{3}{l}{Total time for $\nu_z^{i}$} & $\approx$~107 ms\\
		\hline
		&  &  &\\
		\hline
		\hline
		\multicolumn{4}{c}{Modified cyclotron frequency $\nu_+^{i}$} \\ 
		\hline
		\hline
		Step & & Trap  & Time \\
		\hline
		I) & Pulse excitation at $\nu_+^{i}$ &  PT  & < 1 ms \\
		II) & Free-precession time &  PT  & 100 ms \\  
		III) & Pulse excitation at $\nu_+^{i}$ &  PT  & < 1 ms \\ 
		IV) & $\pi$-pulse at $\nu_+ - \nu_z$  &  PT  & 1 ms \\
		V) & Proton and Be Transport & -  & 1 ms\\
		VI) & Coulomb coupling & CT  & 4 ms\\
		VII) & Proton and Be Transport & - & 1 ms\\
		VIII) & Be detection & BT & 1 ms\\
		\hline
		\multicolumn{3}{l}{Total time for $\nu_+^{i}$} & $\approx$~108 ms\\
		\hline
	\end{tabular}
	\label{table:1}
\end{table}

For a complete $g$-factor measurement, the axial and modified cyclotron frequencies will be determined by scanning external excitation frequencies $\nu_z^{i}$ and $\nu_+^{i}$ close to $\nu_z$ and $\nu_+$, respectively. The cyclotron frequency can be calculated by using Eq.~\ref{inv_the}, where $\nu_- = \nu_z^2 / 2\nu_+$~\cite{brown_precision_1982}. As before, the Ramsey method can be used to measure these motional frequencies by using pulses with duration of a few tens of $\mu$s~\cite{wolf_motional_2019}. In addition, for the modified cyclotron frequency, a pulse excitation at $\nu_+ - \nu_z$ must be applied in order to couple the  modified cyclotron and axial modes. The duration of this pulse can be inferred from Ref.~\cite{cornell_mode_1990}. It is important to note that the cooling of any of the ion motions can be performed at any moment of the complete experimental process without a significant increase in the measurement time.

Systematic shifts of the ion frequencies based on energy- and time-dependent variations in the precision trap~\cite{ketter_first-order_2014}, where the spectroscopy takes place, as well as the frequency stability of the ion transitions will be the fundamental sources of uncertainty. Since the \ppbar will be in a well-defined energy state close to the ground state, systematic corrections based on high-order electric- and magnetic-field imperfections as well as relativistic effects will contribute with fractional uncertainties at the sub-ppt level since they are related to the energy of the ion eigenstates~\cite{vogel_trapped_2010, major_charged_2006}. In the case of the image charge effects, uncertainties due to this correction at the ppt level are expected~\cite{schneider_double-trap_2017}. However, our detection method would allow the use of larger trap sizes, which would reduce this error to levels around 1 ppt~\cite{porto_series_2001}. Time-dependent variations of the electric and magnetic field can be reduced to the sub-ppt levels by using linear interpolations of several axial-frequency measurements~\cite{schneider_double-trap_2017} as well as several measurements of the qubit transition of the Be ion (see Sec.~\ref{Ground-state cooling}) to interpolate the magnetic-field variations. 

Assuming that frequency uncertainties of the ion transitions at a level higher than 1 ppt are dominated by the frequency stability of the ion transitions, under some considerations~\cite{ludlow_optical_2015}, the stability for fractional frequency deviations $y$ will be given by the quantum projection noise~\cite{itano_quantum_1993} 
\begin{equation}
	\label{noise}
	\sigma_{y} = \frac{1}{2\pi\nu_0\sqrt{t_{f}\tau}},
\end{equation}
where $\nu_0$ is the transition frequency and $\tau$ the total measurement duration.
In the case of the Larmor transition, for a magnetic field of $5\,$T, a stability of $2\times10^{-9}/\sqrt{\tau/\mathrm{s}}$ is estimated. For the axial and modified cyclotron frequencies, the stabilities are $5\times10^{-7}/\sqrt{\tau/\mathrm{s}}$ and $7\times10^{-9}/\sqrt{\tau/\mathrm{s}}$, respectively. This implies an expected stability at the 10 ppt level for the Larmor frequency after measurement times of around 12 hours. For the cyclotron frequency, the expected stability would increase by a factor of about 5.5 for the same measurement time, which should be divided equally between the axial and modified cyclotron frequencies. The magnetron frequency would contribute at the sub-ppt level by using the relation $\nu_- = \nu_z^2 / 2\nu_+$. It should be noted, that the motional frequencies are not a two-level system and a factor $1/2|\alpha|$~\cite{wolf_motional_2019} in the frecuency stabilities is expected. Here we assume $|\alpha|\approx1$ for the lowest sensitivity to field-inhomogeneity induced systematics. In addition, frequency stability degradation (see, e.g., Ref.~\cite{dick_local_1987}) due to dead time on the measurement cycle is presumed to be well below the quantum projection noise instability~\cite{ludlow_optical_2015, santarelli_frequency_1998}. Therefore, uncertainties at the order of several tens of ppt in the \ppbar $g$-factor would be expected with a measurement duration on a sidereal-time scale.

\begin{table}[t]
	\caption{Constraints on various $\tilde{b}_p^{J}$, $\tilde{b}_p^{*J}$, $\tilde{b}_{F,p}^{JK}$, and $\tilde{b}_{F,p}^{*JK}$ SME coefficients. The second column contains estimates derived in Ref.~\cite{ding_lorentz_2019} from two recent BASE measurements~\cite{schneider_double-trap_2017,smorra_parts-per-billion_2017}. The third column lists constraints projected to be achievable with the methods discussed in this work.}
	\centering
	\begin{tabular}{ c l l }
		\hline
		\hline
		Coefficient &  Estimated constraints   &  Projected constraints  \\
		\hline
		\hline
		$\lvert\tilde{b}_p^{Z}\rvert$ & $< 8 \times 10^{-25} \mathrm{GeV}$   & $< 1 \times 10^{-26} \mathrm{GeV}$ \\  
		$\lvert\tilde{b}_p^{*Z}\rvert$ &  $< 1 \times 10^{-24} \mathrm{GeV}$   & $< 1 \times 10^{-26} \mathrm{GeV}$ \\
		$\lvert\tilde{b}_{F,p}^{XX} + \tilde{b}_{F,p}^{YY}\rvert$ &  $< 4 \times 10^{-9} \mathrm{GeV^{-1}}$   & $< 4 \times 10^{-11} \mathrm{GeV^{-1}}$ \\
		$\lvert\tilde{b}_{F,p}^{*XX} + \tilde{b}_{F,p}^{*YY}\rvert$ & $< 3 \times 10^{-9} \mathrm{GeV^{-1}}$   & $< 4 \times 10^{-11} \mathrm{GeV^{-1}}$ \\
		$\lvert\tilde{b}_{F,p}^{ZZ}\rvert$ & $< 3 \times 10^{-9} \mathrm{GeV^{-1}}$   & $< 2 \times 10^{-11} \mathrm{GeV^{-1}}$ \\
		$\lvert\tilde{b}_{F,p}^{*ZZ}\rvert$ & $< 1 \times 10^{-8} \mathrm{GeV^{-1}}$   & $< 2 \times 10^{-11} \mathrm{GeV^{-1}}$ \\
		\hline
		$\lvert\tilde{b}_p^{X}\rvert$ &   & $< 1 \times 10^{-25} \mathrm{GeV}$ \\  
		$\lvert\tilde{b}_p^{*X}\rvert$ &     & $< 1 \times 10^{-25} \mathrm{GeV}$ \\
		$\lvert\tilde{b}_p^{Y}\rvert$ &    & $< 1 \times 10^{-25} \mathrm{GeV}$ \\  
		$\lvert\tilde{b}_p^{*Y}\rvert$ &     & $< 1 \times 10^{-25} \mathrm{GeV}$ \\
		$\lvert\tilde{b}_{F,p}^{XZ}\rvert$ &    & $< 1 \times 10^{-10} \mathrm{GeV^{-1}}$ \\
		$\lvert\tilde{b}_{F,p}^{*XZ}\rvert$ &    & $< 1 \times 10^{-10} \mathrm{GeV^{-1}}$ \\
		$\lvert\tilde{b}_{F,p}^{YZ}\rvert$ &    & $< 1 \times 10^{-10} \mathrm{GeV^{-1}}$ \\
		$\lvert\tilde{b}_{F,p}^{*YZ}\rvert$ &    & $< 1 \times 10^{-10} \mathrm{GeV^{-1}}$ \\		
		$\lvert\tilde{b}_{F,p}^{XX} - \tilde{b}_{F,p}^{YY}\rvert$ &     & $< 4 \times 10^{-10} \mathrm{GeV^{-1}}$ \\
		$\lvert\tilde{b}_{F,p}^{*XX} - \tilde{b}_{F,p}^{*YY}\rvert$ &    & $< 4 \times 10^{-10} \mathrm{GeV^{-1}}$ \\
		$\lvert\tilde{b}_{F,p}^{XY}\rvert$ &    & $< 2 \times 10^{-10} \mathrm{GeV^{-1}}$ \\
		$\lvert\tilde{b}_{F,p}^{*XY}\rvert$ &    & $< 2 \times 10^{-10} \mathrm{GeV^{-1}}$ \\
		\hline
	\end{tabular}
	\label{table:2}
\end{table}

These rapid measurements of the \ppbar frequencies would allow access to those SME coefficients discussed in Sec.~\ref{SME implications} that are associated with sidereal variations,
which remain experimentally unexplored in current $g$-factor proton experiments~\cite{ding_lorentz-violating_2016, ding_lorentz_2019}. Following Ref.~\cite{ding_lorentz-violating_2016} and applying the techniques discussed in this work to a hypothetical experiment located at BASE CERN with a colatitude of 43.8$^{\circ}$ and a magnetic-field strength of $5\,$T pointing to the local zenith, several bounds on SME coefficients given by
\begin{eqnarray}
\label{bound_1}
\nonumber\Big\lvert \Delta\tilde{b}_p^{Z} + (3 \times 10^{-16} \mathrm{GeV}^2)(\Delta\tilde{b}_{F,p}^{XX} + \Delta\tilde{b}_{F,p}^{YY}) \\
+ (6 \times 10^{-16} \mathrm{GeV}^2)\Delta\tilde{b}_{F,p}^{ZZ} \Big\rvert \lsim 1 \times 10^{-26} \mathrm{GeV}
\end{eqnarray}
for constant coefficients, 
\begin{eqnarray}
\label{bound_2}
\nonumber&\bigg( \Big[0.7\Delta\tilde{b}_p^{X} + (9 \times 10^{-16} \mathrm{GeV}^2)\Delta\tilde{b}_{F,p}^{XZ}\Big]^2\\
\nonumber& + \Big[0.7\Delta\tilde{b}_p^{Y} + (9 \times 10^{-16} \mathrm{GeV}^2)\Delta\tilde{b}_{F,p}^{YZ}\Big]^2 \bigg)^{1/2}\\
 &\lsim 9 \times 10^{-26} \mathrm{GeV}
\end{eqnarray}
for the first harmonic of sidereal variations and 
\begin{eqnarray}
\label{bound_3}
\nonumber&\bigg( \Big[(2 \times 10^{-16} \mathrm{GeV}^2)(\Delta\tilde{b}_{F,p}^{XX} - \Delta\tilde{b}_{F,p}^{YY}) \Big]^2 \\
\nonumber&+ \Big[(4 \times 10^{-16} \mathrm{GeV}^2)\Delta\tilde{b}_{F,p}^{XY}\Big]^2 \bigg)^{1/2}\\ 
&\lsim 9 \times 10^{-26} \mathrm{GeV}
\end{eqnarray}
for the second harmonic of sidereal variations, respectively, are obtained. Here, $\Delta\tilde{b}_p^{J} \equiv (\tilde{b}_p^{J} - \tilde{b}_p^{*J})/2$ and $\Delta\tilde{b}_{F,p}^{JK} \equiv (\tilde{b}_{F,p}^{JK} - \tilde{b}_{F,p}^{*JK})/2$ with $J, K = X, Y, Z$. In addition, a fractional precision of the antiproton and proton magnetic moments of 100 and 10 ppt are estimated for measurement of sidereal variations and constant effects, respectively. The resulting constraints on these tilde coefficients are summarized in Table~\ref{table:2}. Natural units with $c = \hbar = 1$ are used. The upper part of Table~\ref{table:2} shows the constant coefficients, which are measured with current techniques, and the lower part displays the previously unexplored coefficients in the SME's proton sector that could be accessed with the techniques presented in this work.

\section{Summary}
\label{Summary and outlook}

CPT violation is closely intertwined with a breakdown of Lorentz symmetry,
so that sidereal modulations of physical observables 
in a terrestrial laboratory 
represent a key experimental signature 
for such effects. 
To exploit this insight for investigations in Penning traps, 
the temporal resolution for measurements 
should be at subsidereal time scales, 
a criterion difficult to meet in existing experimental setups involving the \ppbar. 
We have identified the complexity and duration of the currently used spin-readout techniques
as the primary target for substantial improvements. 

More specifically, 
we have argued that recent experimental advances 
in laser-based quantum logic spectroscopy 
have opened an avenue for shortening high-fidelity (anti-)proton spin readout 
by three orders of magnitude.
This idea can be realized via a co-trapped $^9$Be$^+$ `logic' ion:
a SWAP operation
essentially permits the conversion of the (anti-)proton's spin-state information 
to motional excitations of the Be$^+$ ion,
which are readily detectable 
with conventional methods. 
 
In this context, 
we have presented the key features of the experimental implementation of this idea.
In particular,
we have described the design of the required cryogenic multitrap setup 
consisting of interconnected laser, coupling, sideband, and precision traps,
supplemented by an antiproton storage trap. 
We have further detailed the associated experimental protocol
with particular focus on the physical underpinnings of the SWAP operation. 
Finally, 
we have provided estimates for time- and error-budgets. 
These considerations have allowed us to project
improvements by up to two orders of magnitude 
for existing constraints
as well as sensitivity to previously unmeasured effects 
tripling the number of experimentally accessible CPT-violating SME coefficients
at mass dimension up to six.

\begin{acknowledgments}
We are grateful for discussions with D.J.\ Wineland, J.J.\ Bollinger, R.C.\ Thompson, and P.O.\ Schmidt. This work was supported by PTB, LUH, and DFG through the clusters of excellence QUEST and QuantumFrontiers as well as through the Collaborative Research Center 1227 (DQ-mat) and ERC StG ``QLEDS.'' We also acknowledge financial support from the RIKEN Pioneering Project Funding and the MPG/RIKEN/PTB Center for Time, Constants and Fundamental Symmetries. R.L.\ acknowledges support by the Alexander von Humboldt Foundation.
\end{acknowledgments}

\selectlanguage{english}

\bibliography{qls_method}

\end{document}